\documentclass[prl,twocolumn,showpacs,superscriptaddress]{revtex4}
\usepackage{epsfig}
\usepackage{amsmath,amssymb}

\newcommand{\unit}[1]{\ensuremath{\,\mathrm{#1}}}

\newcommand{\state}[3]{\ensuremath{{^{#1}{\mathrm{#2}}_{#3}}}}

\newcommand{\ket}[1]{\ensuremath{|#1\rangle}}

\newcommand{\Pm}{\ensuremath{P_{|\psi^-\rangle}}}
\renewcommand{\P}{\ensuremath{\mathcal{P}}}

\begin{document}

\title{A heralded quantum gate between remote quantum memories}
\author{P.~Maunz}
\email{pmaunz@umd.edu}
\author{S.~Olmschenk}
\author{D.~Hayes}
\author{D.~N.~Matsukevich}
\affiliation{Joint Quantum Institute and Department of Physics, University of Maryland, College Park, MD  20742}
\author{L.-M.~Duan}
\affiliation{FOCUS Center and Department of Physics, University of Michigan, Ann Arbor, MI 48109}
\author{C.~Monroe}
\affiliation{Joint Quantum Institute and Department of Physics, University of Maryland, College Park, MD  20742}

\begin{abstract}
  We demonstrate a probabilistic entangling quantum gate between two
  distant trapped ytterbium ions. The gate is implemented between the
  hyperfine ``clock'' state atomic qubits and mediated by the
  interference of two emitted photons carrying frequency encoded
  qubits. Heralded by the coincidence detection of these two photons,
  the gate has an average fidelity of $90\pm 2 \%$.  This entangling gate
  together with single qubit operations is sufficient to generate
  large entangled cluster states for scalable quantum computing.
\end{abstract}
\maketitle

The conventional model of quantum computing, the quantum circuit
model~\cite{deutsch_quantum_1989, blatt_entangled_2008}, consists of
unitary quantum gate operations followed by measurements at the end of
the computation process to read out the result.  An equivalent model
of quantum computation, which may prove easier to implement, is the
``one-way'' quantum computer~\cite{ briegel_persistent_2001,
  raussendorf_one-way_2001, raussendorf_computational_2002}, where a
highly entangled state of a large collection of qubits is prepared and
local operations and projective measurements complete the quantum
computation.

Experiments with entangled photon states have demonstrated basic
quantum operations~\cite{walther_experimental_2005,
  lu_experimental_2007} for one-way quantum computation. However,
these experiments did not use quantum memories, and the photonic
cluster states used as the resource for the computation are based on
postselection and cannot easily be scaled~\cite{bodiya_scalable_2006}.
In contrast, large entangled states of quantum memories can be
generated using a photon-mediated quantum gate where the number of
necessary operations asymptotically scales linearly with the number of
nodes~\cite{duan_efficient_2005, duan_probabilistic_2006,
  barrett_efficient_2005}.  The successful operation of the gate is
heralded by the coincidence detection of two photons.  Because the
entangling gate is mediated by photons, it can in principle be applied
to a wide variety of quantum memories such as trapped ions, neutral
atoms in cavities, atomic ensembles or quantum dots.

\begin{figure}
\epsfig{file=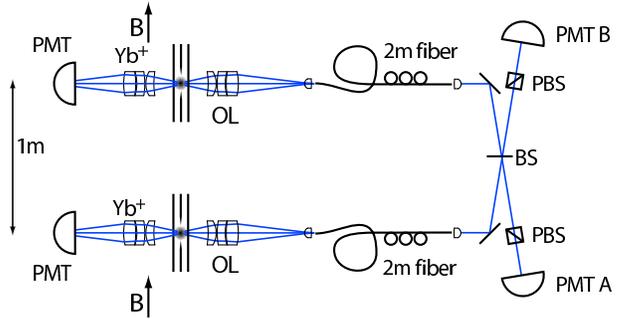,width=\columnwidth}
\caption{The experimental apparatus. Two $^{171}$Yb$^+$ ions are
  trapped in identically constructed ion traps separated by one
  meter. A magnetic field $B$ is applied perpendicular to the
  excitation and observation axes to define the quantization
  axis. About $2\%$ of the emitted light from each ion is collected by
  objective lenses (OL) with numerical aperture of $0.23$ and coupled
  into two single-mode fibers. Polarization control paddles are used
  to adjust the fiber to maintain linear polarization. The output of
  these fibers is directed to interfere on a polarization independent
  $50\%$ beamsplitter. Polarizers (PBS) transmit only the
  $\pi$-polarized light from the ions. The photons are detected by
  single-photon counting photomultiplier tubes (PMT A and PMT
  B). Detection of the atomic state is done independently for the two
  traps with dedicated photomultiplier tubes (PMTs).  }
\label{setup}
\end{figure}
In this Letter, we demonstrate this probabilistic, heralded entangling
gate for two ytterbium ions confined in two independent traps
separated by one meter. The gate is implemented between the long-lived
hyperfine ``clock'' states and mediated by photons carrying frequency
encoded qubits. Unlike the recent demonstration of teleportation
between two ions~\cite{olmschenk_quantum_2009}, here we demonstrate
and characterize the gate for arbitrary quantum states of both qubits,
as required for scalable quantum computing.  We perform the gate on a
full set of input states for both qubits and measure an average
fidelity of $90\pm 2 \%$.  For the particular case that should result
in the antisymmetric Bell state, we perform full tomography of the
final state.

The gate has many favorable properties. First, the ions do not have to
be localized to the Lamb-Dicke regime and the operation is not
interferometrically sensitive to the optical path length
difference. Because the qubits are encoded in the atomic hyperfine
``clock'' states and two well-separated photonic frequency states the
system is highly insensitive to external influences.  Finally, the
operation of the gate between remote ions facilitates individual
addressing for single qubit operations and measurement and there is no
need to shuttle ions.  While the success probability of the gate in
the current experiment is very small ($2.2 \times 10^{-8}$), the
scaling to large quantum networks is still efficient (polynomial
instead of exponential)~\cite{duan_efficient_2005,
  duan_probabilistic_2006}, furthermore, it should be possible to
significantly improve this rate for practical applications.

\begin{figure}
\epsfig{file=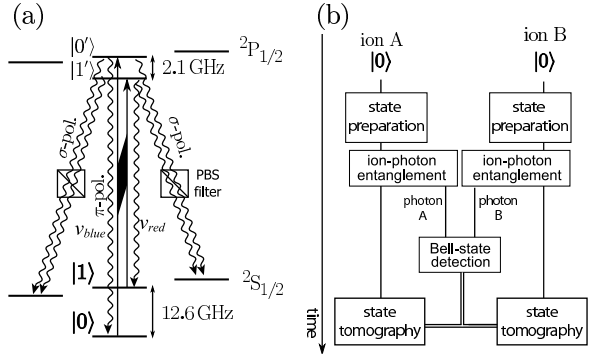,width=0.95\columnwidth}
\caption{(a) Level and excitation scheme. The qubit is encoded in the
  $\state{2}{S}{1/2}$ hyperfine ``clock'' states of the $^{171}$Yb$^+$
  ion and is coherently excited to the $\state{2}{P}{1/2}$ hyperfine
  excited states by a pulse from an ultrafast laser with a wavelength
  centered near $369.5\unit{nm}$. Upon spontaneous emission of a
  $\pi$-polarized photon, the frequency state of the photon is
  entangled with the qubit state of the atom. A polarizer (PBS) blocks
  photons from different decay channels. (b) Gate operation
  scheme. After initialization in $\ket{0}$, ion $1$ and $2$ are
  prepared in the input states $\ket{\Psi_a}_1$ and $\ket{\Psi_a}_2$,
  respectively. The frequency of each spontaneously emitted
  $\pi$-polarized photon is entangled with the state of the respective
  ion. If these two photons are detected in the antisymmetric Bell
  state, the quantum state of the two ions is projected on the state
  $\ket{\Psi_{aa}} \propto Z_1(I-Z_1Z_2)\ket{\Psi_a}_1\ket{\Psi_a}_2
  $. Here $Z_i$ is the Pauli-$z$ operator acting on ion $i$.}
\label{entanglement_scheme}
\end{figure}

%
We trap two single $^{171}$Yb$^+$ atoms in two identically constructed
Paul traps, located in independent vacuum chambers separated by
approximately one meter (Fig.~\ref{setup}). 
An ion typically remains in the trap for several
weeks. Doppler-cooling by laser light slightly red-detuned from the
$\state{2}{S}{1/2} \leftrightarrow \state{2}{P}{1/2} $ transition at
$369.5\unit{nm}$ localizes the ions to better than the diffraction
limit of the imaging system but not to the Lamb-Dicke regime. With a
probability of about $0.5\%$, the excited $\state{2}{P}{1/2}$ state
decays to the meta-stable $\state{2}{D}{3/2}$ level. This level is
depopulated with a laser near $935.2\unit{nm}$ to maintain efficient
cooling and state detection. We apply an external magnetic field
$B=5.2\unit{Gauss}$ to provide a quantization axis, break the
degeneracy of the atomic states, and suppress coherent dark state
trapping.
The atomic qubit is encoded in two $\state{2}{S}{1/2}$ ground-state
hyperfine levels of the $^{171}$Yb$^+$ atom, with $\ket{0} :=
\ket{F=0,m_F=0}$ and $\ket{1} := \ket{F=1,m_F=0}$, which have a 
separation of $12.6\unit{GHz}$ (Fig.~\ref{entanglement_scheme}
(a)). Here $F$ is the total angular momentum of the ion and $m_F$ its
projection along the quantization axis. These hyperfine ``clock''
states are to first order insensitive to the magnetic field and thus
form an excellent quantum memory~\cite{fisk_accurate_1997,
  olmschenk_manipulation_2007}.

The remote gate protocol is shown schematically in
Fig.~\ref{entanglement_scheme}. We first initialize each ion in
$\ket{0}$ with a $1\unit{\mu s}$ pulse of light resonant with the
$\state{2}{S}{1/2} (F=1) \leftrightarrow \state{2}{P}{1/2} (F=1)$
transition. Then we independently prepare each ion ($i=1,2$) in any
desired superposition state $\ket{\Psi_a}_i = \alpha_i\ket{0}_i +
\beta_i\ket{1}_i$ by applying a resonant microwave pulse with
controlled phase and duration $(0-16\unit{\mu s})$ directly to one of
the electrodes of each trap.
Next, we use an ultrafast $\pi$-polarized resonant laser pulse to
simultaneously transfer the superposition from the ground state qubit
states to the $\state{2}{P}{1/2}$ hyperfine states $\ket{0^\prime} :=
\ket{F^\prime=1, m_F^\prime=0}$ and $\ket{1^\prime} :=
\ket{F^\prime=0, m_F^\prime=0}$ of each ion with near-unit
efficiency. For $\pi$-polarized light the dipole selection rules allow
only the transitions $\ket{0} \leftrightarrow \ket{0^\prime}$ and
$\ket{1} \leftrightarrow \ket{1^{\prime}}$ which have equal transition
strength. The two transitions are well-resolved, as their center
frequencies are separated by $\Delta\nu = 14.7\unit{GHz}$, while the
natural linewidth of the excited state is only about
$20\unit{MHz}$. The bandwidth of the $1\unit{ps}$ pulse of about
$300\unit{GHz}$ is broad compared to $\Delta\nu$ but small compared to
the fine structure splitting in Yb$^+$ of about $100\unit{THz}$,
allowing both transitions to be driven equally while the population of
the excited $\state{2}{P}{3/2}$ state remains vanishingly
small. Consequently, the qubit can be transferred coherently from the
\state{2}{S}{1/2} ground state to the excited \state{2}{P}{1/2}
state~\cite{madsen_ultrafast_2006}.

Following excitation, each ion will emit a single photon. Upon
emission of a $\pi$-polarized $369.5\unit{nm}$ photon, the
frequency-mode of the emitted photon and the state of the ion are in
the entangled state
\begin{equation}
  \ket{\Psi_{ap}}_i = \alpha_i\ket{0}_i\ket{\nu_b}_i +
  \beta_i\ket{1}_i\ket{\nu_r}_i,
\end{equation} 
where $\ket{\nu_b}$ and $\ket{\nu_r}$ are the two possible frequency
states of the emitted photon. The state of the total system is
$\ket{\Psi_{apap}} = \ket{\Psi_{ap}}_1 \otimes \ket{\Psi_{ap}}_2$. For
each ion, emitted photons are collected with a lens with a numerical
aperture of $0.23$ and are coupled into a single-mode fiber. The
output of the fiber from each ion is directed to interfere on a
polarization-independent $50\%$ beamsplitter. Each output of the
beamsplitter is directed through a linear polarizer and detected with
a single-photon counting photomultiplier tube.
To ensure high contrast interference of the two photons from different
ions the photons must be indistinguishable. To this end, we first
carefully minimize the micromotion of the ions to prevent modulation
of the emission frequency. Second, the geometrical mode from the two
fibers is matched to better than $98\%$ as characterized with laser
light. Third, the emitted photons are matched in their arrival time at
the beamsplitter to better than $100\unit{ps}$. As a consequence of
the quantum interference~\cite{hong_measurement_1987, shih_new_1988,
  braunstein_measurement_1995}, two photons, each in a superposition
of two frequency modes, can only emerge from different output ports of
the beamsplitter if they are in the antisymmetric state
$\ket{\psi^-_{pp}} = (\ket{\nu_b}_1\ket{\nu_r}_2 -
\ket{\nu_r}_1\ket{\nu_b}_2)/\sqrt{2}$. Upon a coincidence detection
event the ions are projected onto the state
\begin{align}
  \nonumber\ket{\Psi_{aa}} &= \frac{1}{\sqrt{2 \Pm}} \left(
    \alpha_1\beta_2\ket{0}_1\ket{1}_2 -
    \beta_1\alpha_2\ket{1}_1\ket{0}_2 \right) \\
  &= \frac{1}{\sqrt{2 \Pm}}
  \frac{Z_1(I-Z_1Z_2)}{2}\ket{\Psi_a}_1\ket{\Psi_a}_2,
\end{align}
where $Z$ is the Pauli-$z$ operator and $\Pm =
(\alpha_1^2\beta_2^2+\beta_1^2\alpha_2^2)/2$ is the probability to
find the photons in the antisymmetric Bell state. Thus the coincidence
detection of two photons heralds the operation of the remote two-ion
quantum gate $Z_1(I-Z_1Z_2)$. In contrast to a simple entanglement
process~\cite{simon_robust_2003}, the final state depends on the initial
states of both ions, and hence this scheme must keep track of the
initial states and phases of the two qubits. This property of an
entangling gate is essential for the generation of cluster states of
more than two ions as the entangling gate has to preserve any
entanglement which may already be present. Being a projection or
measurement gate this process is not unitary. Indeed, for the input
states $\ket{0}_1\ket{0}_2$ and $\ket{1}_1\ket{1}_2$ a heralding event
should never occur, which would be calamitous in the circuit
model. However, in the protocol to generate cluster states, the input
qubits are always by design in a superposition of the two qubit
states. In this case, the protocol succeeds with a nonvanishing
probability and scales favorably~\cite{duan_probabilistic_2006}.


\begin{figure}
\epsfig{file=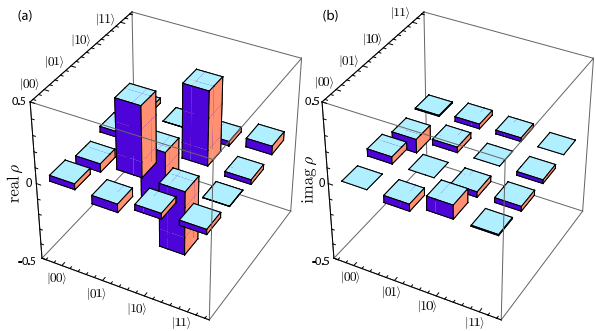,width=.95\columnwidth}
\caption{State Tomography of the final state $\ket{\Psi_{aa}} \propto
  Z_1(I-Z_1Z_2) \left( \ket{0+1}_1\ket{0+1}_2 \right)$. Real (a)
  and imaginary (b) part of the reconstructed density matrix.  The
  fidelity of the expected output state $\ket{0}_1\ket{1}_2 -
  \ket{1}_1\ket{0}_2$ of the gate is $F=0.87(2)$. The generated state has a
  concurrence of $C=0.77(4)$ and an entanglement of formation of
  $E_F=0.69(6)$. The density matrix is obtained with a maximum likelihood
  algorithm from 601 events measured in 9 different
  bases. }
\label{state_tomography}
\end{figure}
To verify the operation of the gate we first characterize the
generation of the maximally entangled antisymmetric Bell state
$\ket{\Psi_{aa}} = (\ket{0}\ket{1} - \ket{1}\ket{0})/\sqrt{2}$ by full
state tomography. Both ions are prepared in
$\ket{\Psi_a}_i = ( \ket{0}_i + \ket{1}_i )/\sqrt{2}$ and operated on
by the gate. We then measure the state of each ion in three 
mutually unbiased bases. Detection of the quantum state of each ion in
the $x$ ($y$) bases is done by first applying a resonant microwave
$\pi/2$-pulse with a relative phase of $0$ ($\pi/2$) with respect to
the initial microwave pulse. Standard fluorescence detection is then
used to determine the quantum state of the ion; the ion scatters light
if found in state \ket{1}, while it remains dark if found in
\ket{0}~\cite{olmschenk_manipulation_2007}. Consequently, we get an
answer in every attempt to measure the state and this answer is
correct with more than $98\%$ probability. The density matrix
(Fig.~\ref{state_tomography}) is obtained using a maximum likelihood
algorithm~\cite{james_measurement_2001}. From this density matrix we
calculate an entangled state fidelity of $F=0.87(2)$, a concurrence of
$C=0.77(4)$ and an entanglement of formation $E_F=0.69(6)$. The
entanglement of this state is considerably higher than in
our previous experiments~\cite{moehring_entanglement_2007,
  matsukevich_bell_2008} due to technical improvements and the
superior coherence properties of the photonic frequency and atomic
``clock''state qubits.

\begin{table*}
  \caption{Results of the remote quantum gate process. Listed in this
    table are the input states of the two ions, the expected
    output state of the gate, the measurement performed, the number of
    heralding events, the obtained fidelity, the measured and ideal
    probability for two photons to be in the antisymmetric Bell
    state. The fidelity of the output states is obtained by parity
    measurements in the appropriate bases. Here, the parity $\P_{xy}$
    is the difference of the probabilities to find the two ions in the
    same state and in opposite states when ion $1$, $2$ is measured in
    the $x$, $y$ basis, respectively. The other parity values are
    defined similarly. From these results we calculate the average
    gate fidelity $\overline{F}=0.90(2)$. The success probability of
    the gate is $P_{\text{gate}} = \Pm \times 8.5\times 10^{-8}$.} 
\label{process_fidelity}
\begin{ruledtabular}
\begin{tabular}{ccccccc}
  input state & expected state & measurement & events & fidelity &
  $\Pm$ (meas.) & \Pm (theo.) \\
  $\ket{0+1}\otimes\ket{0+1}$&$\ket{0}\ket{1}-\ket{1}\ket{0}$&$\frac{1}{4}\left(
    1-\P_{xx}-\P_{yy}-\P_{zz} \right)$ & $210$ & $0.89(2)$ & $0.26(1)$ & $1/4$ \\
  $\ket{0+i}\otimes\ket{0+1}$&$\ket{0}\ket{1}-i\ket{1}\ket{0}$&$\frac{1}{4}\left(
    1-\P_{xy}+\P_{yx}-\P_{zz} \right)$ & $179$ &$0.86(2)$ & $0.26(1)$ & $1/4$ \\
  $\ket{0-1}\otimes\ket{0+1}$&$\ket{0}\ket{1}+\ket{1}\ket{0}$&$\frac{1}{4}\left(
    1+\P_{xx}+\P_{yy}-\P_{zz} \right)$ & $178$ & $0.85(1)$ & $0.22(2)$ & $1/4$ \\
  $\ket{0-i}\otimes\ket{0+1}$&$\ket{0}\ket{1}+i\ket{1}\ket{0}$&$\frac{1}{4}\left(
    1+\P_{xy}-\P_{yx}-\P_{zz} \right)$ & $188$ & $0.81(2)$ & $0.27(2)$ & $1/4$\\
  $\ket{0+1}\otimes\ket{1}$ & $\ket{0}\otimes\ket{1}$ & $\frac12 \left(
    1+\P_{zz} \right) $ & $42$ & $0.86(5)$ & $0.24(4)$ & $1/4$\\
  $\ket{0}\otimes\ket{0+1}$ & $\ket{0}\otimes\ket{1}$ & $\frac12 \left(
    1+\P_{zz} \right) $ & $52$ & $0.90(4)$ & $0.20(3)$ & $1/4$\\
  $\ket{0}\otimes\ket{1}$   & $\ket{0}\otimes\ket{1}$ & $\frac12 \left(
    1+\P_{zz} \right) $ & $48$ & $0.98(2)$ & $0.39(6)$ & $1/2$\\
  $\ket{0}\otimes\ket{0}$   & $0$                     &              &
  $65$ &  & $0.04(1)$ & $0$\\
\end{tabular}
\end{ruledtabular}
\end{table*}

To characterize the functionality of the gate for arbitrary input
states we measure the fidelity of the output state for a
representative set of input states as shown in
Table~\ref{process_fidelity} and obtain an average fidelity of
$\overline{F}=0.90(2)$.

We do not characterize the action of the gate on
certain input states that differ only by global qubit rotations.  Such
states are identical to those considered up to an overall choice of
basis, so the input states listed Table I are representative of a full
set of unbiased qubit bases.

The observed entanglement and gate fidelity are consistent with known
experimental imperfections. The primary error sources that reduce the
gate fidelity are imperfect state detection $(3\%)$, geometrical mode
mismatch on the beamsplitter $(6\%)$, and detection of
$\sigma$-polarized light due to the non-zero solid angle and
misalignment of the magnetic field $(<2\%)$. Micromotion at the
rf-drive frequency of the ion trap, which alters the spectrum of the
emitted photons and can degrade the quantum interference, is expected
to contribute to the overall error less than $1\%$. Other error
sources include imperfect state preparation, pulsed excitation to the
wrong atomic state, dark counts of the PMT leading to false
coincidence events, mismatch of the quantization and polarizer axes,
and multiple excitation due to pulsed laser light leakage, and are
each estimated to contribute much less than $1\%$ to the overall
error.

The entangling gate demonstrated here is a heralded probabilistic
process where the success probability is given by

\begin{equation*}
  P_{\text{gate}} =
  \Pm \left( p_\pi \frac{ \Delta\Omega}{4\pi}
    T_{\text{fiber}} T_{\text{optics}} \eta \right) ^2 \approx  \Pm 2.2
  \times 10^{-8}.
\end{equation*} 

Here $p_{\pi} = 0.5$ is the probability that a collected
$369.5\unit{nm}$ photon is $\pi$-polarized, $\Delta\Omega/4\pi = 0.02$
is the solid angle from which photons are collected,
$T_{\text{fiber}}=0.2$ is the coupling and transmission efficiency of
the collected light through the single-mode fiber,
$T_{\text{optics}}=0.95$ is the transmission coefficient of the other
optics and $\eta=0.15$ is the quantum efficiency of the
photomultiplier tube. The probability to find two photons in the
antisymmetric Bell state $\Pm$ is a function of the initial states and
$0 \le \Pm \le 1/2$. As a coincidence detection of two photons is
necessary to herald the operation of the gate, the success probability
scales as the square of the photon detection probability.

In the current experiment photons are collected by a lens in free
space which only covers a small solid angle. Increasing the collection
solid angle, which can be possible by using parabolic
mirrors~\cite{lindlein_new_2007} or microstructured
lenses~\cite{streed_scalable_2008}, can significantly increase the
success probability. Furthermore, the spontaneous emission into free
space could be replaced by the induced emission into the small mode
volume of a high finesse cavity~\cite{hennrich_vacuum-stimulated_2000,
  mckeever_deterministic_2004} which can reach near unit efficiency.
Even though the free spectral range of the cavity would have to be
$14.7\unit{GHz}$ to simultaneously support both frequency modes,
choosing a near-concentric design could still result in a small
mode-volume and thus in a high emission probability into a
well-defined Gaussian mode. Alternatively, a fast microwave
$\pi$-pulse together with the photonic time-bin encoded qubit could be
used~\cite{barrett_efficient_2005}.  While the success probability in
the current experiment is small, employing the aforementioned
techniques to increase the photon collection probability may
dramatically increase the success probability of the gate and could
make the generation of large entangled cluster states feasible.

We have demonstrated a probabilistic, heralded entanglement gate
between two remote matter qubits with an average fidelity of
$\overline{F}=0.90(2)$. The remote entangling gate demonstrated here
could be used to realize long-distance quantum repeaters and to
demonstrate a loophole-free Bell-inequality violation. Furthermore,
together with local operations, the entangling gate may be used to
scalably generate cluster states for the realization of a one-way
quantum computer~\cite{briegel_persistent_2001,
  raussendorf_one-way_2001, raussendorf_computational_2002,
  duan_efficient_2005, duan_probabilistic_2006}.

\begin{acknowledgments}
  This work is supported by the Intelligence Advanced Research Project
  Activity (IARPA) under Army Research Office contract, the National
  Science Foundation Physics at the Information Frontier program, and
  the NSF Physics Frontier Center at JQI.
\end{acknowledgments}


\end{document}